\documentclass[preprint2]{aastex62}

\def\MSUN{M$_{\sun}$}

\def\kms{$\rm km~s^{-1}$}
\def\simlt{\lower.5ex\hbox{$\; \buildrel < \over \sim \;$}}
\def\simgt{\lower.5ex\hbox{$\; \buildrel > \over \sim \;$}}
\def\vlsr{v_{\rm LSR}}
\def\micron{$\mu$m}
\def\spitzer{{\it Spitzer}}
\newcommand{\feii}{[\ion{Fe}{2}]}

\newcommand{\sili}{[\ion{Si}{1}]}
\newcommand{\silii}{[\ion{Si}{2}]}
\newcommand{\hei}{[\ion{He}{1}]}
\newcommand{\siii}{[\ion{S}{3}]}
\newcommand{\siv}{[\ion{S}{4}]}
\newcommand{\oiv}{[\ion{O}{4}]}

\graphicspath{{./}{figures/}}

\received{}
\revised{}
\accepted{}
\submitjournal{ApJ}

\shorttitle{Unshocked Ejecta in Cas A}
\shortauthors{Raymond et al.}
\begin{document}

\title{The Temperature and Ionization of Unshocked Ejecta in Cas A}

\author[0000-0002-7868-1622]{J.C. Raymond}
\email{jraymond@cfa.harvard.edu}
\affil{Center for Astrophysics, 60 Garden St., Cambridge, MA 02176, USA}

\author[0000-0002-2755-1879]{B.-C. Koo}
\affil{Department of Physics and Astronomy, Seoul National University
Gwanak-gu Seoul 08826, Korea}

\author{Y.-H. Lee}
\affil{Department of Physics and Astronomy, Seoul National University
Gwanak-gu Seoul 08826, Korea}

\author[0000-0002-0763-3885]{D. Milisavljevic}
\affil{Department of Physics and Astronomy, Purdue University, West Lafayette, IN 47907, USA}

\author[0000-0003-3829-2056]{R.A. Fesen}
\affil{6127 Wilder Lab, Department of Physics \& Astronomy, Dartmouth College, Hanover, NH 03755, USA}

\author[0000-0002-7924-3253]{I. Chilingarian}
\affil{Center for Astrophysics, 60 Garden St., Cambridge, MA 02176, USA}
\affiliation{Sternberg Astronomical Institute, M.V.Lomonosov Moscow State University, Universitetsky prospect 13, Moscow, 119992, Russia}



\begin{abstract}

The supernova remnant Cassiopeia A (Cas A) is one of the few remnants in which it is possible to observe unshocked ejecta. A deep 1.64 $\mu$m image of Cas A shows a patch of diffuse emission from unshocked ejecta, as well as brighter emission from Fast-Moving Knots and Quasi-Stationary Flocculi.  Emission at 1.64 $\mu$m is usually interpreted as [Fe II] emission, and spectra of the bright knots confirm this by showing the expected emission in other [Fe II] lines.  We performed NIR spectroscopy on the diffuse emission region and found that the unshocked ejecta emission does not show those lines, but rather the [Si~I] 1.607 $\mu$m line. This means that the 1.64 $\mu$m line from the unshocked ejecta may be the [Si~I] 1.645 line from the same upper level, rather than [Fe~II].  We find that the [Si ~I] line is formed by recombination, and we use the [Si~I] to [Si II] ratio to infer a temperature about 100 K, close to the value assumed for analysis of low frequency radio absorption and that inferred from emission by cool dust. Our results constrain estimates of Cas A's total mass of unshocked ejecta that are extremely sensitive to temperature assumptions, but they do not resolve the ambiguity due to clumping.

\end{abstract}



\section{Introduction} \label{sec:intro}

Cas A is a young core-collapse supernova remnant (SNR).  It was created by a type IIb supernova approximately 340 years ago, and it is about 3.4 kpc away \citep{krause08, reed95}. It has been extensively studied from radio wavelengths through gamma-rays.  In the optical and near IR, the remnant  shows up as the Fast Moving Knots (FMKs) of ejecta mostly devoid of H and He emission moving at thousands of $\rm km~s^{-1}$ and slower Quasi-Stationary Flocculi (QSFs) of circumstellar material enriched in He and N \citep[e.g.]{milisavljevic13, koo13, alarie14, fesen16, lee17}.  Both are dense knots of gas heated by relatively slow shock waves.  The X-ray and radio show synchrotron emission, and the X-rays also show thermal emission from shocked ejecta and circumstellar gas \citep{vinklaming, delaney10, hwanglaming, onicurosevic}.

Cas A is one of the few SNRs in which it is possible to observe the unshocked ejecta.  The cool, freely expanding gas inside the reverse shock has been detected in IR emission lines with the {\it Spitzer} satellite \citep{smith09, delaney10, isensee10}, through low frequency radio absorption \citep{kassim95, arias18}, in the near IR as [S III] emission lines \citep{milisavljevic15} and as a cold component of the infrared dust emission \citep{delooze17}.  It is detectable because of photoionization and heating by the X-ray and EUV emission from gas heated by the forward and reverse shocks.

This paper reports deep IR spectra of diffuse emission inside Cas A obtained with the MMT/Magellan Infrared Spectrograph (MMIRS) instrument on the 6.5 m MMT telescope
\citep{mcleod12}. The diffuse emission appears in a deep 1.64 $\mu$m image from UKIRT \citep{koo18} as a smooth, faint emission region interior to the shell of FMKs.  

While emission at 1.64 $\mu$m is usually identified with the strong [Fe II] line at 1.644 $\mu$m, the [Si I] line at 1.645 $\mu$m is also possible, as has been observed in the nebular phases of supernovae \citep{kjaer10, milisavljevic17}.  In Cas A, the diffuse emission is correlated with [Si II] 35 $\mu$m emission rather than $^{44}$Ti or [O IV]/[Fe II] 25 $\mu$m emission in the {\it Spitzer} data, suggesting the [Si I] identification \citep{koo18}.  The MMIRS spectra can be used to test that
identification.

We compare the inferred ejecta properties with those obtained from other line ratios from {\it Spitzer} and ISO \citep{docenko10} and with the mass and temperature of unshocked dust \citep{delooze17}.

\section{Observations} \label{sec:observations}
%   - MMT, date, positions, PAs, resolutions, ranges
%   - 0.2" pix, 6.9' FOV, 2048 HgCdTe, 0.95-2.45 micron
%   - offset, noise level
%   - Fig. 1. Slit positions overlaid on the [Fe II]+[Si I] image
%?  also show Spitzer [Si II] and [O IV]+[Fe II]

We carried out near-infrared (NIR) spectroscopic observations toward the central region of Cas A using MMIRS at the MMT 6.5-m telescope in November and December, 2016. MMIRS is equipped with a HAWAII-2 $2048\times2048$-pixel HgCdTe detector
with a pixel scale of $0\farcs2$, which provides a
$6.'9\times6.'9$ field-of-view for imaging observations.
We used MMIRS in long-slit spectroscopy mode with 
a $6.'9$-long slit 
utilizing the $J+zJ$ and $H3000+H$ (grism$+$filter) sets  
to obtain $J$-band (0.95--1.5 \micron) and $H$-band (1.50--1.80 \micron) spectra.
The slit width was fixed to $0\farcs6$, so that 
the mean spectral resolutions of the $J$- and $H$-band spectra are  
$R\sim2000$ and $R\sim2800$, respectively.

The slits are centered at the explosion center 
at $(\alpha,\delta)_{\rm J2000}=(23^{\rm h} 23^{\rm m} 27.77^{\rm s}, +58^\circ 48' 49\farcs4)$ 
\citep{thorstensen01} (Figure~\ref{fig-slit}).
In total, $J$- and $H$-band spectra at four slit positions have been obtained.  
At two slit positions, i.e., Slit 1 and Slit 2, 
however, there were offsets of a few arcseconds 
between the $J$- and $H$-band slits. 
For Slit 1, the telescope also drifted during the observation.
These problems imposed some 
limitations on our analysis of the spectra at these slit positions, 
but since the source is extended, 
the spectra still provide useful information. 

For the sky background subtraction, 
we took separate off-source spectra at each slit position. 
The single exposure time was 300 sec,
while the total integration time for individual $J$- and $H$-band spectra   
varied from 10 min to 30 min.
The seeing during the observation was $0\farcs7$--$1\farcs3$.
For the data reduction,
we used the MMIRS pipeline written in the IDL language \citep{chilingarian15}.
The wavelength calibration and distortion correction were done
by using bright OH airglow emission lines falling in the $J$- and $H$-band spectra, while 
the absolute photometric calibration and correction for telluric
absorption was done 
by comparing the observed standard A0V star (HD240290)
with the Kurucz model spectrum of Vega\footnote{\url{http://kurucz.harvard.edu/}}.
We compared the \feii\ 1.644 \micron\ flux in the flux-calibrated spectra
to that in the 1.64 \micron\ narrow-band image \citep{koo18}, and found 
that the uncertainty in the absolute flux scale is 20--30\%.

\section{Results and Analysis}

\subsection{MMIRS Spectra}
 
%   - Fig. 2. J and H band spectra  \\
%   - lack of 1.257 mu, presence of other [Si I] \\
%   - implies that emission in the [Fe II]+[Si I] image is due to [Si I] line. \\

The results are summarized in Table 1 where the observed intensities, Doppler velocities and line widths 
of the \feii\ 1.257 \micron\ and/or \feii/\sili\ 1.64 \micron\ 
lines toward the emission features covered by individual slits in Figure 1 are listed. Examples of 2D spectra are shown in Figure 2.
For undetected lines, the upper limits ($3\sigma$) are given. Note that some emission 
features are covered only by either $J$- or $H$-band slit, so that the table is not complete.  
In Table 1, we also give the line fluxes 
estimated from the deep \feii+\sili\ 1.64 \micron\ narrow-band image for comparison.  

The \feii\ 1.257 \micron\ flux 
assumes that the flux of the narrow-band image is due to 
\feii\ 1.644 \micron\ line and that   
the intrinsic intensity ratio of \feii\ 1.257 \micron\ to \feii\ 1.644 \micron\ line is 1.36 
\citep{deb11}. Note that this intrinsic ratio is uncertain by 20\%  
\citep[see references in][]{koo16}.
In scaling the 1.64 \micron\ flux of the narrow band image to the \feii\ 1.257 \micron\ flux, 
the difference in interstellar extinction has been counted, i.e.,    
the H column density toward Cas A varies considerably 
\citep[1.3--2.8$\times10^{22}$ cm$^{-2}$;][]{hwanglaming}, 
so that \feii\ 1.257~\micron\ line experiences 
0.7--1.5 mag of more extinction according to the dust opacity law of the interstellar 
interstellar medium \citep{draine03}.
In the following, we summarize the results for individual emission features. 

\noindent
(1) Slit 1-$J$ crosses the lower parts of interior diffuse clumps IDC 1 and IDC 2,  but no emission lines are detected in either source. Previous \spitzer\ and ground-based 
NIR observations detected high-velocity, broad  \silii\ 34.8 \micron\ and 
\siii\ 906.9/953.1 nm lines from these sources, indicating that they are SN ejecta material 
\citep{delaney10,milisavljevic15}.
If we assume FWHM of 300~\kms, 
the $3\sigma$ upper limit of the \feii\ 1.257~\micron\ line flux would be   
$2.7\times10^{-16}$ erg s$^{-1}$ cm$^{-2}$,
which is slightly less than    
the flux we expect from the deep 1.64 \micron\ narrow-band image.
Therefore, our result suggests that  
the emission in the 1.64 \micron\ narrow image is {\em probably} due to the 
\sili\ 1.645 \micron\ line, although we couldn't confirm it because 
Slit 1-$H$ barely missed the sources (Figure \ref{fig-slit}).

\noindent
(2) Slit 2-$J$ crosses the southern diffuse emission (SDE),
and we detected a weak emission feature at 10\arcsec\ above 
the main ejecta shell bright in \feii\ 1.257~\micron\ emission.
The location coincides with a bright and compact clump embedded in the SDE in 
Figure \ref{fig-slit}. The line is not resolved and 
its central velocity is $\vlsr\sim -50$~\kms\ which is close to the systemic 
velocity of Cas A, i.e., $\simlt -48$~\kms 
\citep{reynoso97, kilpatrick14}. The low central velocity together with 
the narrow line width suggests that the emission is probably associated with  
the CSM not SN ejecta. 
In contrast to the shocked dense CSM observed in Cas A \citep{lee17}, however, 
it does not show \hei\ 1.083~\micron\ emission.

\noindent
(3) Slit 2-$H$ crosses the western part of IDC 4 (IDC 4-W), and we obtained 
an $H$-band spectrum at its intensity peak position.
The spectrum shows a strong narrow emission line around 1.64~\micron.
We also detected two additional emission lines at 1.60~\micron\ and 1.67~\micron\
corresponding to the counterparts of the \feii\ 1.644~\micron\ line,
indicating that the line is  
\feii\ 1.644~\micron\, not \sili\ 1.645~\micron .
Similar to the above detection in SDE, the line is not resolved and its 
central velocity is $\vlsr\approx -56$~\kms. 
These line parameters imply that
the clump-like feature embedded in IDC 4 is probably CSM.
(This, however, does not mean that the diffuse source IDC 4 is CSM. See below.) 
The \feii\ line ratios are \feii\ 1.677/\feii\ $1.644\sim0.07~(0.02)$ and
\feii\ 1.600/\feii\ $1.644\sim0.05~(0.01)$, implying  
electron densities of $\sim 3000$ cm$^{-3}$ according to CHIANTI version 8 \citep{delzanna15}.
This density is an order of magnitude lower than
that of the shocked QSF knots \citep[i.e., well above 5000 cm$^{-3}$;][]{lee17}.
%BCK Check density
%SDE

\noindent
(4) Slits 3-$J$ and -$H$ cross the `pillar' which is a 
filament protruding from the southwestern main ejecta shell. 
No emission lines are detected in the $J$ band.
The $3\sigma$ upper limit of \feii\ 1.257~\micron\ line 
assuming $\Delta v_{\rm FWHM}=150$~\kms\ is 
$\sim 3$ times larger than the flux expected from the narrow-band image, so that 
the non-detection does not rule out the \feii\ line origin for the pillar.
In the $H$ band, we clearly detected narrow emission lines 
at 1.64 \micron. We also barely detected emission lines of similar velocity structure 
at 1.61~\micron\ corresponding to the position of the \sili\ companion line.
Their flux ratio is consistent with the intrinsic ratio given by  
the ratio of their Einstein $A$ 
coefficients, which is $(7.14\times 10^{-4})/(2.01\times10^{-3})\approx 1/3$.
This strongly suggests that the pillar is due to 
the \sili\ line. The central velocity of \sili\ 1.645 \micron\ emission line 
decreases from $-1400$~\kms\ at the bright southern end  
near the main ejecta shell
to $-1600$~\kms\ at the northern faint end of the pillar.
The high central velocity and the small velocity width indicate that 
the pillar is unshocked Si SN ejecta. The unshocked ejecta interpretation is supported by shock models that predict very high [Fe II] to [Si I] ratios and observations of strong Fe emission in both FMKs and QSFs \citep{koo18}.

\noindent
(5) Slit 4 crosses the eastern part of IDC 4 (IDC 4-E).
In $J$ band, we have not detected any emission line.
The $3\sigma$ upper limit to the \feii\ 1.257~\micron\ line is 
$\sim 3$ times larger than the flux estimated from
the narrow band image.
In the $H$ band, 
we detected very weak and broad emission features 
composed of three velocity components
around 1.64~\micron, which could be either
\feii\ 1.644~\micron\ or \sili\ 1.645~\micron\ emission lines.
Their radial velocities are about 
$-3000$, $-1000$, and $+1000$~\kms,
and the line widths seem to be broader than $300$~\kms. 
These line parameters suggest that 
the diffuse source IDC 4-E is probably SN ejecta. It also suggests that 
the connected diffuse source IDC 4-W, the morphology of which is similar to 
IDC 4-E, is likely SN ejecta too.   

To summarize, 

\noindent
(1) For IDC 1 and IDC 2, which are SN unshocked ejecta material with Si and S, we have not detected \feii\ 1.257 \micron\ line. So their emission in the 1.64 \micron\ narrow image is 
probably due to the \sili\ 1.645 \micron\ line. 
(2) For SDE and IDC 4-W, we detected low-velocity, narrow \feii\ lines from embedded 
bright clumps, which are most likely the CSM clumps. 
The electron densities implied by the \feii\ lines is a few $10^3$~cm$^{-3}$. 
The nature of the extended diffuse emission is uncertain. 
(3) For the pillar in the southwestern area, 
we detected \sili\ 1.645 \micron\ and 1.607 \micron\ lines 
with central velocities of $\vlsr=-1400$~\kms\ to $-1600$~\kms,  
indicating that it is unshocked Si SN ejecta. 
(4) For IDC 4-E, we detected broad emission lines at 1.64 \micron\  
with central velocity of $-3000$~\kms\ to $+1000$~\kms, indicating that 
it is unshocked SN ejecta material. But because of the limited sensitivity of the $J$-band spectrum, 
we could not conclude whether the emission is due to \feii\ or [Si~I].
This result for IDC 4-E suggests that the diffuse source IDC 4-W is also probably SN ejecta 
material.

\subsection{\spitzer\ Mid-Infrared Line Fluxes}

\begin{figure}
\includegraphics[width=\textwidth]{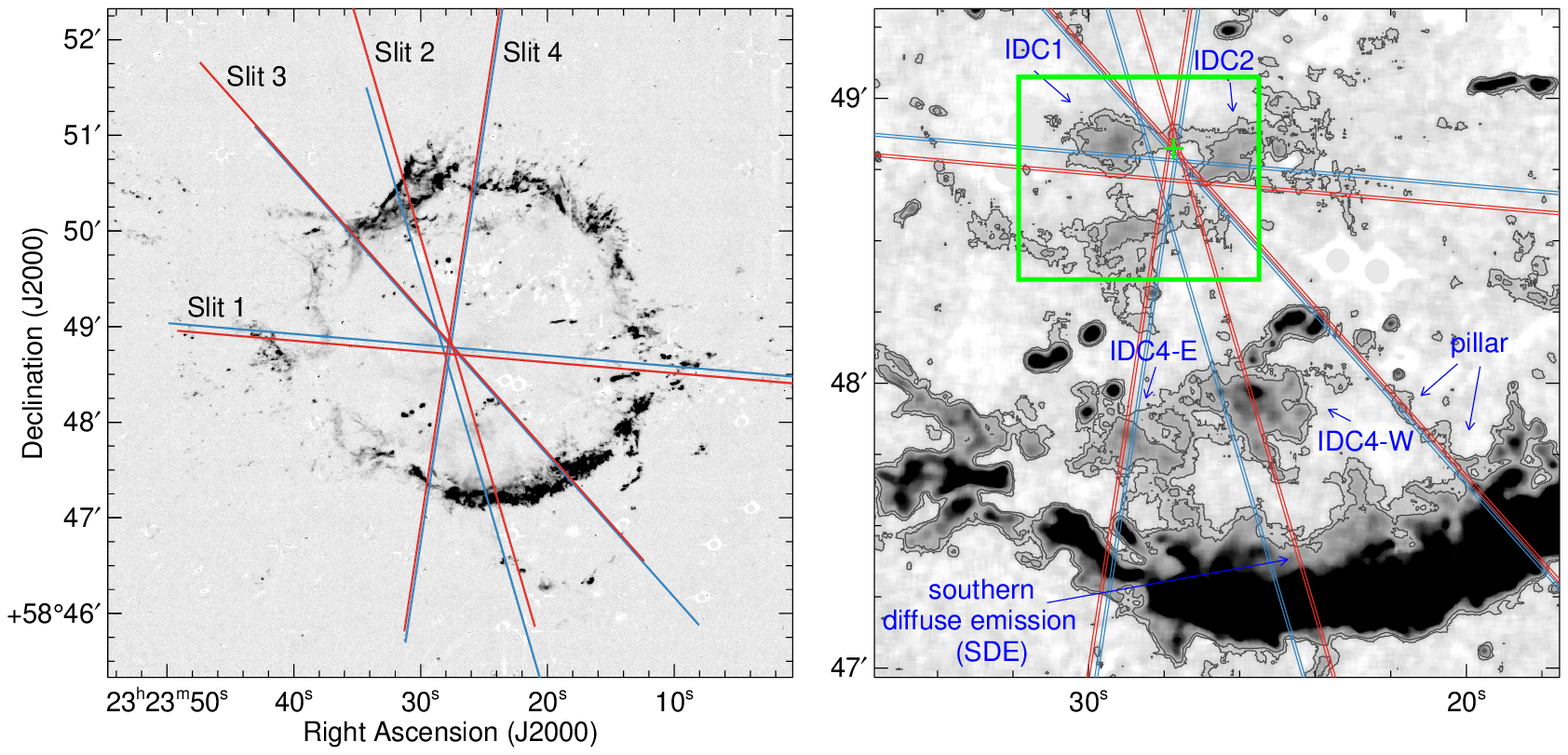}
\caption{Left: Slit positions of the MMIRS long-slit spectroscopic observation of Cas A. 
	The blue and red bars represent J- and H-band slit positions, respectively. 
    The background image is the deep \feii+\sili\ 1.64 \micron\ image of Cas A \citep{koo18}. 
    Right: An enlarged view of the slit positions in the central area. The background image is the same 1.64 \micron\ image but with contours. The green cross marks the explosion center \citep{thorstensen01}, 
%BCK the figure has been replaced and the % following sentence has been added.
while the green box represents the area where the MIR line fluxes in Table 2 are obtained. Some prominent emission features are labeled following \cite{koo18} 
  and discussed in the text. 
} \label{fig-slit}
\end{figure}

Our MMIRS results in \S~3.1 suggest that the central diffuse emission in Figure~\ref{fig-slit} is {\em probably} \sili\ 1.645 \micron\ emission.  In this section, we derive \spitzer\ mid-infrared (MIR) line fluxes to be compared with the \sili\ 1.645 \micron\ flux. The entire area of Cas A had been mapped by the \spitzer\ Space Telescope  with two Infrared Spectrograph (IRS) low-resolution modules (R = 60-130) 
SL (5--15 \micron)  and LL (15--38 \micron) \citep{smith09}, 
while the central area surrounding the explosion center 
had been also mapped with two high-resolution ($R\sim 600$) modules   
SH (10--20 \micron) and LH (20--35 \micron) \citep{isensee10,isensee12}. 
The data were downloaded from the \spitzer\ Heritage Archive\footnote{http://sha.ipac.caltech.edu/applications/Spitzer/SHA/}. 
The deep 1.64 \micron\ narrow-band image in Figure~\ref{fig-slit} was obtained with a filter centered at 1.645 \micron\ and an effective bandwidth of 0.0284 \micron\ (or 5200~\kms in radial velocity), so we use the high-resolution data cubes to obtain the flux covering the same velocity range. The pixel scales of the SH and LH modules are  $2\farcs3$ and $4\farcs5$, respectively. The low-resolution module data have been used to estimate the possible contribution from the background emission.

Since the mapping area of the IRS SH module is smaller than
that of the IRS LH module, we extract  the average profiles 
from the area of the IRS SH module, i.e., the rectangular area 
marked by the green box in the left frame of Figure \ref{fig-slit}.
Figure \ref{fig-spitzer} (a) shows the extracted profiles 
of the \siv\ 10.51 \micron, \siii\ 18.71 \micron, \oiv+\feii\ 25.91 \micron, \siii\ 33.48 \micron,
and \silii\ 34.81 \micron\ lines. All lines have similar profiles.

We have derived line fluxes between $v=-2600$ and +2600~\kms\ from these average 
profiles and they are given in Table \ref{table-flux}. The fastest unshocked ejecta move at nearly 5000 \kms , so we definitely miss some emission near the center of the ejecta.  Based on the [Si II] velocity structure in Figure 3 (see also Figure 4 of Smith et al. 2009), the total flux may be about 20\% higher. For the \silii\ 34.81 \micron\ flux, 
there is a contribution from the background emission, but we estimate that it 
is negligible ($\simlt 2$\%) and we have not subtracted it out. In the table, 
the \sili\ 1.645 \micron\ flux 
is the corresponding flux derived from the 
deep \feii+\sili\ image assuming that 
all the emission is \sili. The quoted uncertainty includes the absolute flux calibration uncertainty.

The table also shows the fluxes corrected for the extinction assuming $N_{\rm H}=1.9\times 10^{22}$~cm$^{-2}$ (or $A_V$=10 mag) which is the mean absorbing H column density toward the IRS SH area from X-ray analysis \citep{hwanglaming}.
The measured extinction to the central region of Cas A, however, is uncertain, ranging from 6 mag to 11 mag (see Table 3 of \citealt{koo17} and references therein), so, for example, if $A_V=6$ mag, the extinction-corrected 
\sili\ 1.645 \micron\ flux would be 
$\sim 1/2$ of the flux in Table 3.

Figure \ref{fig-spitzer} shows the spectral line maps obtained by integrating 
over this velocity range. 
It is clear that the morphology of the 
\silii\ 34.81 \micron\ emission is very similar to that of the \feii+\sili\ 1.64 \micron\ emission. 
The S line emission also shows some similar emission features, e.g., the bright IDC1,  
while the \oiv\ line emission has  a morphology quite different from the other lines.    
It is worthwhile to mention that these line emissions including the 
\feii+\sili\ 1.64 \micron\ emission have a diffuse, spatially-extended 
component filling the mapping area, and the fluxes in Table \ref{table-flux} 
include these diffuse components.

\subsection{Analysis}

%We must first identify the line that dominates the diffuse emission in Figure XX.  Normally it would be taken to be the [Fe II] $a^4D_{7/2}~-~a^6F_{9/2}$ 1.644 $\mu$m line, but it turned out to be [Si I] $^3P_2~-~^1D_2$ line at 1.645 $\mu$m in supernovae \citep{kjaer10, milisavljevic17}. We identify the diffuse emission in Cas A with the [Si I] line because 1) the emission is well correlated with the [Si II] 35$\mu$m emission but not the [O IV] + [Fe II] 25 $\mu$m emission in Figure XX, and 2) the spectrum of the diffuse region does not show any emission in the [Fe II] 1.259 $\mu$m line, which is intrinsically brighter than the 1.644 $\mu$m line, but it does show a weak feature at 1.607 $\mu$m, where the companion [Si I] $^3P_1~-~^1D_2$ line should be seen at 1/3 the brightness of the [Si I] 1.645 $\mu$m line.

To identify the emission mechanism of \sili\ and \silii\ lines, we use the collisional excitation cross section from \citet{pindzola77} and the recombination rate from \citet{abdel-naby12}.  The collisional excitation rate has a Boltzmann factor $e^{(-9060/T)}$, and because Si I is neutral, the cross section approaches zero at threshold instead of maintaining a more constant value.  Both of these mean that the excitation rate is very small at temperatures below a few thousand K, and the [Si I] emission could only be explained  by collisional excitation for unreasonable combinations of temperature and ionization state.  On the other hand, the recombination rate includes very low temperature dielectronic recombination due to resonances near threshold \citep{nahar95, abdel-naby12}.  

We computed the ratio of [Si II] to [Si I] fluxes by assuming that 1/4 of the recombinations of Si$^+$ go by way of the singlet levels of Si$^0$ and cascade through the $^1D_2$ state, while 3/4 go through the triplets.  That is appropriate if most of the recombinations go through high $l$ levels, because the cascade will favor the  $^1D_2$ state.  Dielectronic recombination at very low densities will favor moderate $l$ levels, in which case more of the cascades go through the $^1S_0$ state, and a factor 0.21 would be more accurate, but we do not know which states contribute to the recombination.  
We use the collision strength for the [Si II] 35 $\mu$m line from \citet{tayal08}. The result is independent of the ionization fractions, because both lines are formed by collisions of electrons and Si$^+$ ions, and it is shown in Figure~\ref{si2_si1_ratio}. The [Si II] to [Si I] line ratio of the central diffuse emission in Cas A is $\sim 90$ (Table 2), so we obtain a temperature of $\sim 100$ K. If the extinction is 6 mag in $A_V$ \citep[e.g.,][]{eriksen09}, the line ratio would be 180 so that the temperature becomes 120 K. Because the ratio is extremely sensitive to temperature, the uncertainty in reddening, has a small effect on the derived temperature.

We can carry this one step farther by comparing the [Si I] and [Si II] emission at 100 K with the emission measure derived from low frequency radio absorption \citep{arias18}.  Their average value, $37 ~(T/100)^{3/2}~\rm pc~cm^{-6}$, seems to match the emission measure in the region discussed above (their Figure 3), and at 100 K it requires that $n_{Si II}/n_e$ = 0.015 to match the [Si I] and [Si II] intensities.  Since we are missing some emission at high Doppler velocities, the Si$^+$ fraction would be higher by perhaps 20\%.  The small Si II fraction would be due to a combination of the $\rm Si^+$ ionization fraction and the elemental fraction of Si in the ejecta.  However that value is rather sensitive to temperature.  The low temperature also matches the higher density photoionization models of \citet{eriksenphd}, indicating that substantial clumping is present.

The presence of [S III] and [O IV] (Table 2) suggests that much of the silicon may be in Si III and Si IV, but these higher ionization states may be located in warmer gas that does not contribute much to the emission measure.  The temperature T=100 K is an average temperature, and it is not necessarily unique because there could be a contribution to either line from higher temperature gas. However, higher temperature gas would have difficulty accounting for the low frequency radio absorption because of the $T^{-3/2}$ factor in the free-free opacity and the already large amount of mass required at T=100 K \citep{arias18}.  We also note that the morphology of the diffuse [O IV] emission is somewhat different from that of the [Si II]
(Figure 3).  

It is also worthwhile to mention the lack of the [S I] lines at 1.082 and 1.131 $\mu$m that are analogous to the [Si I] lines in the MMIRS spectra.  The [S I] lines are more sensitive to reddening, of course, but S$^+$ may simply lack the strong very low temperature dielectronic recombination that Si$^+$ has. We have not been able to find detailed calculations for the S$^+$ recombination rate.

We can also obtain an estimate for the amount of iron.  The contribution of [Fe II] to the blend of [O IV] and [Fe II] in the diffuse gas is uncertain, but based on the discussions in \citet{isensee10} and \citet{docenko10}, it seems unlikely that [Fe II] contributes more than 30\% to the 25 $\mu$m blend.  Therefore the intensity ratio of [Fe  II] to [Si II] is less than 0.25.  With collision strengths from Version 8 of CHIANTI and a temperature of 100 K, that the number of Fe$^+$ ions is less than 7 times the number of Si$^+$ ions.  We do not have a reliable enough model for the ionization states of those elements to determine a total abundance ratio.

\section{Discussion} \label{sec:discussion}

\subsection{Comparison with Previous Results}

There is some warmer gas present in the unshocked ejecta.  \citet{isensee10} give fluxes of the blue- and red-shifted components of the [S III] lines at 18.7 $\mu$m and 33.5 $\mu$m from a small region in the diffuse emission area, and the ratios correspond to a temperature and density of 1250 K and 450 $\rm cm^{-3}$, respectively, using CHIANTI version 8 \citep{delzanna15} with collision rates from \citet{hudson12}.  \citet{isensee10} quote a wide range of ratios, but most cluster around 0.05, which would imply 250 K.  The ratio of fluxes of 0.5 from a large part of the diffuse emission shown in Table 2 implies a temperature of about 2500 K if the density is 20 $\rm cm^{-3}$, or a temperature several times smaller if the density is above 100 $\rm cm^{-3}$.  

In addition, while \citet{docenko10} concentrate on emission from the FMKs, their position \#1 seems to overlap with our diffuse emission region.  The emission there is faint, and the line ratios are different than in the other regions they study.  The ratio of the ISO [O III] line fluxes at 51.81 $\mu$m and 88.36 $\mu$m from position \#1 in their Table 3 is 0.72, which is close to the low density limit, indicating a density below 100 $\rm cm^{-3}$ according the the atomic data in Version 8 of the CHIANTI database \citep{delzanna15}.  This contrasts with the FMK positions of \citet{docenko10}, where the flux ratio is 2.5 and the density is 500 to 1000 $\rm cm^{-3}$.

\subsection{Comparison with dust mass}

Our temperature supports the assumption of around 100 K used by \citet{arias18} to determine the emission measure of the unshocked ejecta, and therefore their mass estimate of about 3 \MSUN .  That is an upper limit assuming no clumping.  Unfortunately, the silicon IR lines have the same sensitivity to clumping as does the low frequency radio absorption, so we cannot further narrow down the mass.  Indeed, since the photoionization models of \citet{eriksenphd} show low temperatures and strong [Si II] emission at high densities, both the low frequency radio absorption and the IR observations are strongly biased toward high density clumps.  Since both are proportional to density squared, they overestimate the mass by 1/f, where f is the volume filling factor.  On the other hand, they do not account for the mass in the remaining 1-f of the volume. 

\citet{delooze17} estimated a dust mass of 0.4 to 0.6 \MSUN .  That implies a dust formation efficiency of around 10\% unless the unshocked ejecta are mostly clumped into high density regions of order 200 $\rm cm^{-3}$, as is sometimes invoked for the preshock density of the FMKs and suggested by the \citet{eriksenphd} photoionization models.  The clumping indicated by the high densities needed for the observed low ionization states suggests an even higher dust formation efficiency.  However, mass estimates generally indicate that the FMKs account for a small fraction of the ejecta mass.

\subsection{Comparison with simulations}

The mass and distribution of metal-rich supernova ejecta provide key constraints for 3D simulations modeling SN explosions and their subsequent evolution to the remnant phase \citep{janka12, orlando16}. Presently, pressing discrepancies exist between model predictions and observations. One significant issue regards the kinematic distribution of the metal-rich debris. Simulations show that the mass density should essentially be unaffected, and that although mixing can affect the species distribution, the bulk of the heaviest mass including Ni should have velocities below 4000 \kms\ \citep{ono13}. This is, in fact, in stark contrast with what we currently understand of Cas A. Its X-ray-bright Fe (which traces the original $^{56}$Ni distribution and is believed to account for the majority of Fe-rich ejecta) has velocities around and above 4000 \kms\  \citep{hughes00, willingale02, delaney10}.

This theory-observation disconnect implies that simulations are not yet adequately following the dynamics of mixing and/or ``missing iron" remains to be detected in Cas A's interior \citep{milisavljevic15}. Detailed understanding of the progenitor star’s interior structure and its influence on explosion dynamics is an active area of investigation \citep{wongwathanarat17}. Another area of interest is to develop a robust estimate of the total mass of unshocked ejecta. To date, estimates range from $< 0.4~M_\odot$ \citep{hwanglaming, delaney14}, to a few solar masses \citep{arias18}.  Part of the wide range in estimates of unshocked ejecta mass can be attributed to their sensitivity to temperature and density assumptions. Here we have contributed additional constraints on the temperature of the interior gas, but the extent of clumping remains a wide, gaping unknown. As discussed above, both the low frequency radio absorption and our [Si I]/[Si II] estimates refer to the coolest part of the ejecta.  They probably overestimate the mass of that component by the clumping factor, but miss the lower density, warmer and more ionized gas that fills the rest of the volume.

\subsection{Ionization state}

The ejecta are very cold and neutral early on, but they are photoionized and heated by X-rays from the ejecta and circumstellar gas that pass through the reverse and forward shocks, respectively, by EUV emission from the FMKs and QSFs, and by EUV photons from the thin ionization zones just behind the reverse and forward shocks \citep{hamilton88}.  The time scale for photoionization by EUV photons from the ionization zone of the non-radiative reverse shock \citep{hamilton88} is roughly 30 to 300 years.  Photoionization by X-rays is much slower, and photoionization by EUV photons from the FMKs is very uncertain, but is likely to be somewhat slower based on the models of \citet{sutherland95} and the [O III] optical flux of \citet{bevan17}.  

If the photoionization time scale is 100 years and each ionization deposits 10 eV, the heating rate is $1 \times 10^{-19}~\rm erg~cm^{-3}~s^{-1}$ at a density of 20 $\rm cm^{-3}$.  That compares with a cooling rate of $6 \times 10^{-19}~\rm erg~cm^{-3}~s^{-1}$ at that density and 100 K for $\rm Si^+$, but most of the Si is in higher ionization states which are ineffective coolants at low temperatures.  Given the large uncertainties, it is likely that photoionization by UV from the reverse shock maintains both the ionization state and temperature of the unshocked ejecta.  Close to the reverse shocks in the FMKs, both the density and ionization rate may be much larger \citep{docenko10}.

Models by \citet{orlando16} predict a size scale for unshocked ejecta like that observed, though the predicted density seems to be too small.  They do not include photoionization, so it is not possible to compare with the ionization state and temperature that we observe.

\section{Summary}

We find that the 1.64 $\mu$m emission from the unshocked ejecta in Cas A is probably dominated by the [Si I] line formed by recombination and that the temperature is about 100 K.  This is in line with the temperature assumed in the analysis of low frequency radio absorption \citep{arias18}, and it is consistent within large uncertainties with the photoionization and heating rates.  The uncertainties in the temperature estimate are dominated by the reddening correction for the 1.64 $\mu$m line and our plausible, but unverified, assumption that 1/4 of the Si$^+$ recombinations go through the $^1$D upper level of the 1.64 $\mu$m line.

We do not expect the [Si I] line to dominate the fluxes in narrow band 1.64 $\mu$m images in most circumstances because Fe$^+$ is generally more abundant than Si$^0$ at temperatures high enough to excite these lines and because the Fe line is most strongly excited.  Indeed, the 1.64 $\mu$m images of FMKs and QSFs in Cas A are dominated by the [Fe II] line \citep{koo18}, and it is only in cold, ionized gas such as unshocked SN ejecta that [Si I] can compete.

\acknowledgments
We thank the referee for exceptionally helpful comments, along with Tom Gorczycka for discussions of the dielectronic recombination of Si$^+$.  We also thank the staff of the MMT observatory for their help in making the observations. B.-C. K. was supported by Basic Science Research Program through the National Research Foundation of Korea(NRF) funded by the Ministry of Science, ICT and future Planning (2017R1A2A2A05001337).  IC's research is supported by the SAO Telescope Data Center. IC acknowledges the Russian Science Foundation grant 17-72-20119.

\vspace{5mm}
\facilities{MMT(MMIRS), UKIRT}

\clearpage

\clearpage
\begin{figure}
\center
\includegraphics[scale=0.8]{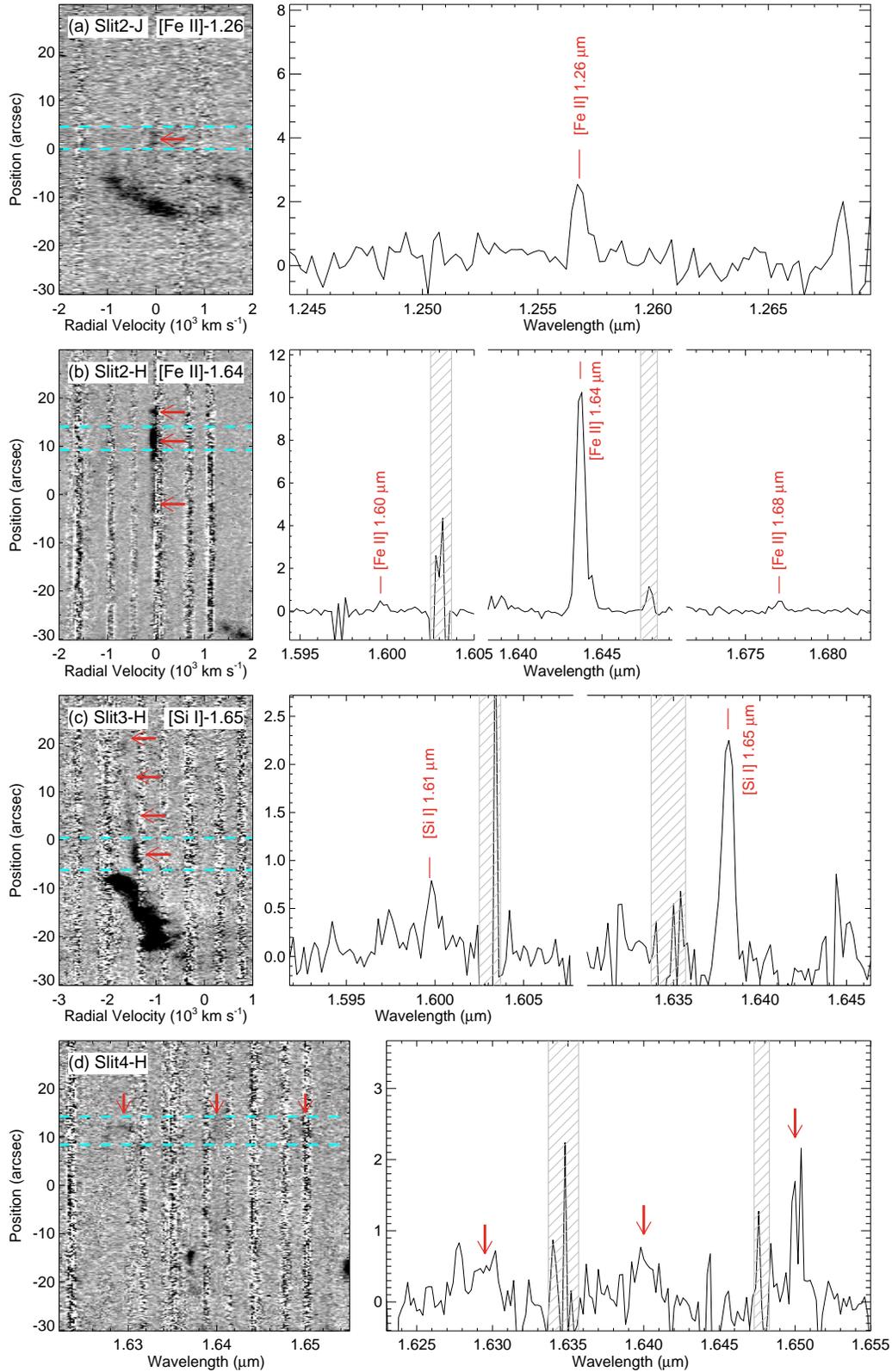}
\caption{Spectra of detected lines. Grey-scale images are two-dimensional 
dispersed slit images, where red arrows mark the detected emission features.
The cyan dashed lines mark the area where 
the one-dimensional spectra are extracted. 
In the one-dimensional spectra, the hatched columns represent the wavelength range contaminated by telluric OH lines, and the 
unit of the $y$ axis is $10^{-13}$ erg s$^{-1}$ cm$^{-2}$ \micron$^{-1}$.
} \label{fig-pv}
\end{figure}

\clearpage
\begin{figure}
\center
\includegraphics[scale=0.7]{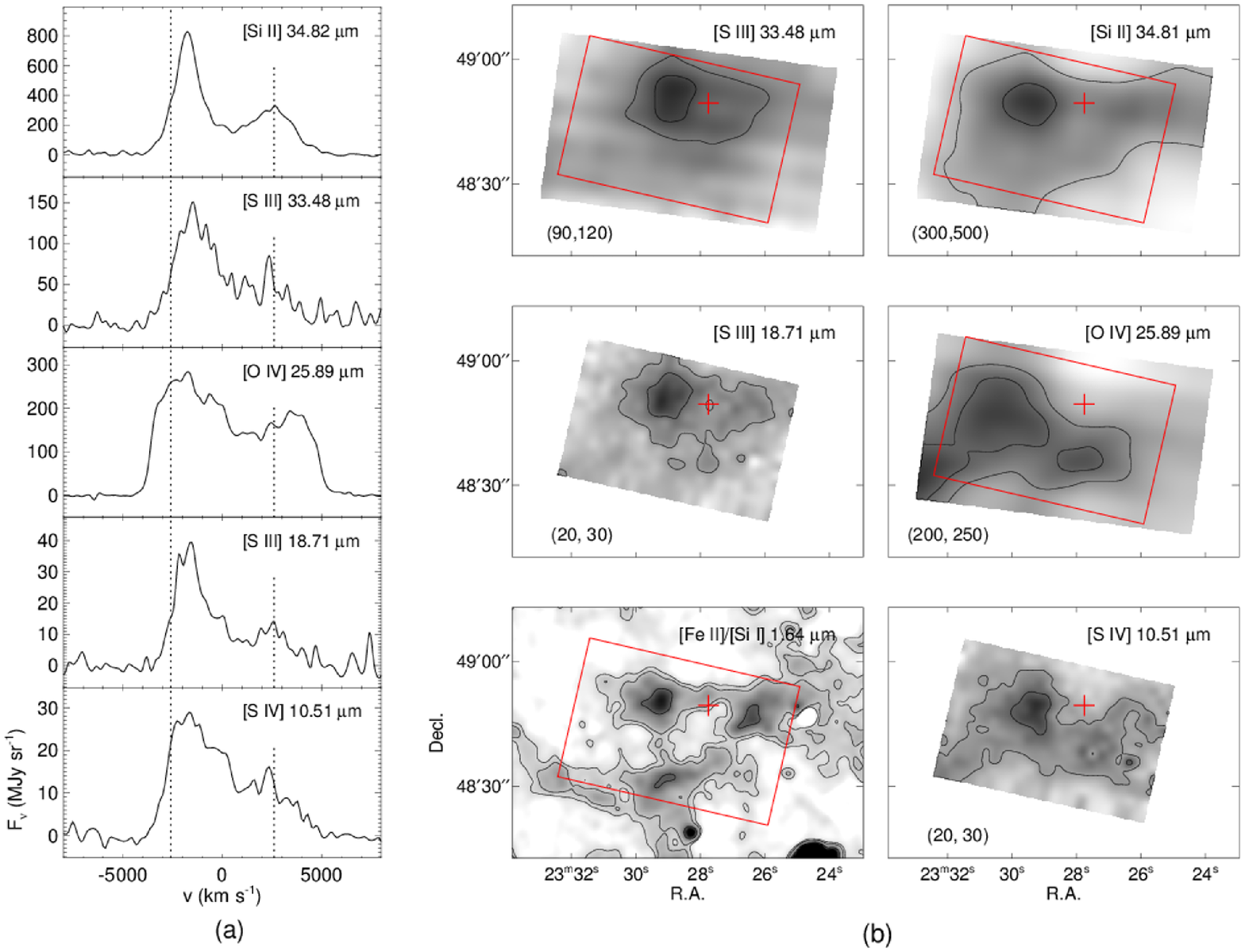}
\caption{(a) Spectra of \spitzer\ MIR ionic lines. 
These are the average spectra of the \spitzer\ IRS SH area, i.e., 
the green box in Figure~\ref{fig-slit} or the red box in (b).  
The dotted lines mark the velocity coverage  
($\pm 2600$~\kms) of the deep \feii+\sili\ 1.64 \micron\ narrow-band image.
(b) Spectral line maps of the central area.  
The \feii+\silii\ 1.64 \micron\ map is from our deep narrow-band image while the MIR line 
maps are produced from the \spitzer\ IRS SH and LH data cubes by integrating over 
$v=-2600$ to +2600~\kms. 
The grey intensity scale is linear.
The contour levels given in the bottom left corners of the \spitzer\ images 
are mean intensity levels in MJy sr$^{-1}$. 
The red box represents the \spitzer\ IRS SH mapping area, while  
the red cross marks the explosion center. 
} \label{fig-spitzer}
\end{figure}

\clearpage
\begin{figure}
\plotone{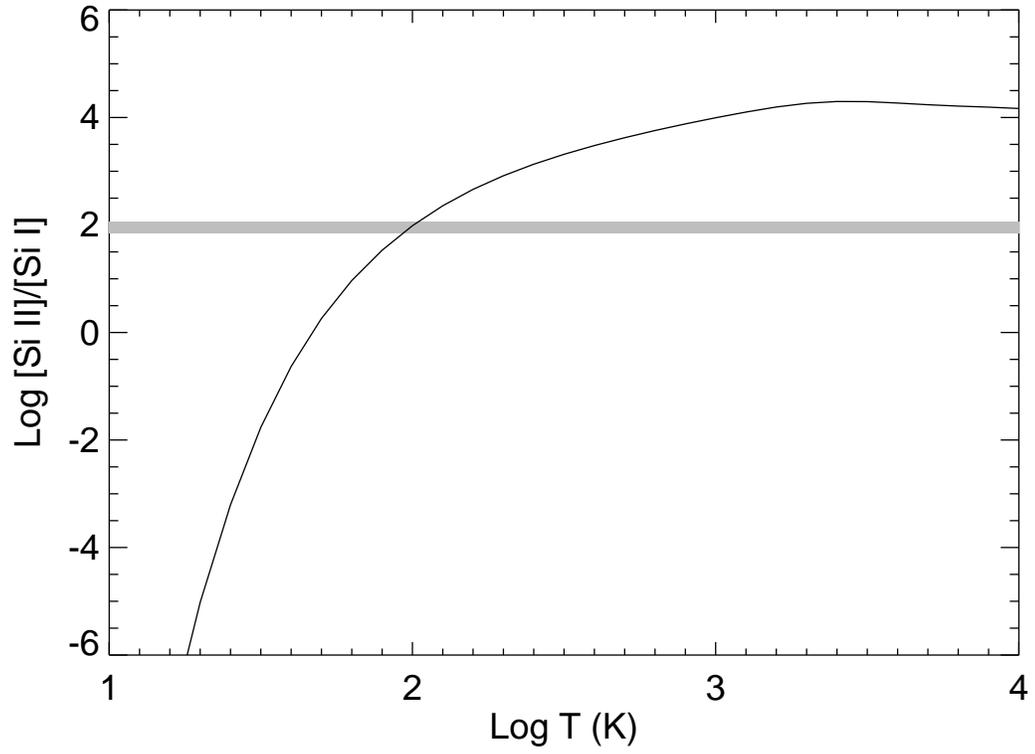}
\caption{Predicted energy flux ratio of [Si II] 35 $\mu$m to [Si I] 1.645 $\mu$m as a function of temperature.  The observed value is shown as a horizontal bar.}
\label{si2_si1_ratio}
\end{figure}


%\clearpage

\clearpage
\begin{deluxetable}{llcccccccl}
\tabletypesize{\scriptsize}
\tablewidth{0pt}
\tablecolumns{10}
\tablecaption{Observed parameters of \feii/\sili\ emission lines \label{tab-flux}}
\tablehead{
 \noalign{\vskip 0.4cm}
 \colhead{} & \colhead{} &\colhead{} 	&\colhead{} 	&
\multicolumn{2}{c}{\feii\ 1.257 \micron}	& \colhead{} &   
\multicolumn{2}{c}{\feii/\sili\ 1.64 \micron}	& \\
\cline{5-6} \cline{8-9}
\colhead{} & \colhead{Source}	& \colhead{$v_{\rm rad}$}  &  \colhead{$\Delta v_{\rm FWHM}^{\rm a}$}	&
\colhead{Spec.}	&	\colhead{Image$^{\rm b}$}	& \colhead{}   &
\colhead{Spec.}	&	\colhead{Image$^{\rm c}$}	&
\colhead{} \\
 \noalign{\vskip -0.1cm}
\colhead{Slit}& \colhead{Name}& \colhead{(\kms)} & \colhead{(\kms)}	& 
\multicolumn{2}{c}{(10$^{-16}$ erg s$^{-1}$ cm$^{-2})$} & \colhead{}   &
\multicolumn{2}{c}{(10$^{-16}$ erg s$^{-1}$ cm$^{-2})$} &
\colhead{Note}}
\startdata
Slit 1-J & IDC 1		&	\nodata &	\nodata 	&
$<2.7^{\rm d}$	&	3.5(0.2)	&  &\nodata 	&	6.4(0.4)	&	Si Ejecta?	\\
Slit 1-J & IDC 2		&	\nodata 	&	\nodata 	&
$<2.7^{\rm d}$	&	3.3(0.2)	&  &	\nodata & 6.4(0.4) 	&	Si Ejecta?	\\
Slit 2-J & SDE			&	$-50 (7)$	&	$153(17)$	&
1.8(0.3)	&	2.1(0.1)	&  &	\nodata &3.7(0.2)	&	CSM$^{\rm e}$	\\
Slit 2-H & IDC 4-W	&	$-56 (2)$	&	$101(2)$	&
\nodata &	3.7(0.2)	&  &	6.6 (0.2)	&	6.4(0.4)	&	CSM$^{\rm e}$	\\
Slit 3-JH & pillar	&	$-1421 (4)$	&	$137(8)$	&	
$<3.4^{\rm d}$ & 1.0 (0.1)  & 	&	1.9(0.1)	&	1.8(0.1)	&	Si Ejecta	\\
Slit 4-JH & IDC 4-E		&	$-3000$ to $+1000$	&	200--360	&
$<5.7^{\rm d}$			&	1.8(0.1)	&  &	3.5(0.6)	&	3.0(0.2)	&	Fe/Si Ejecta?	\\
\enddata
\tablenotetext{a}{Measured FWHM of the line. Spectral resolutions are 
150~\kms\ and 110~\kms\ in $J$- and $H$-bands, respectively.}
\tablenotetext{b}{Expected \feii\ 1.257~\micron\ flux estimated from
the 1.64~\micron\ narrow-band image
assuming that the latter emission is due to \feii\ 1.644 \micron\ emission
(see text for more details).}
\tablenotetext{c}{\feii/\sili\ 1.64~\micron\ flux in the slit estimated from
the 1.64~\micron\ narrow-band image. The slit lengths used to derive the flux are 
$12\farcs4$, $12\farcs4$, $2\farcs8$, $3\farcs0$, $4\farcs8$, and $4\farcs0$
from top (IDC 1) to bottom (IDC 4-E).}
\tablenotetext{d}{$3\sigma$ upper limit. 
For IDC 1 and IDC 2 where \feii/\sili\ 1.64 \micron\ lines are not detected, 
we assumed $\Delta v_{\rm FWHM}=300$~\kms.}
\tablenotetext{e}{This conclusion applies to the bright clumps embedded in the source (see Figure \ref{fig-slit}). 
The nature of the surrounding diffuse emission is uncertain.}
\tablecomments{The symbol ``\nodata'' means that the source was not covered by the slit in 
$J$ (\feii\ 1.257 \micron) or $H$ band (\feii/\sili\ 1.64 \micron).}
\end{deluxetable}

%BCK I have replaced the table.
%\clearpage
%\begin{deluxetable}{lcc}
%\tablecaption{Diffuse Emission Line Fluxes\tablenotemark{a} \label{tab:table}}
%\tablehead{
%\colhead{Ion} & \colhead{Wavelength} & \colhead{Flux} \\
%}
%\startdata
%$[\rm Si I]^b$  & 1.645 $\mu$m  &  0.027 \\
%$[\rm S III]$     & 18.71 $\mu$m  &  0.34  \\
%$[\rm O IV]$      & 25.91 $\mu$m  &  2.40  \\
%$[\rm S III]$    & 33.48 $\mu$m  &  0.72  \\
%$[\rm Si II]$     & 34.81 $\mu$m  &  3.10  \\ 
%\enddata
%\tablenotetext{a}{From 30" Radius region}
%\tablenotetext{b}{Corrected for reddening $A_V$=7.0}
%\end{deluxetable}

\clearpage
\begin{deluxetable}{lcc}
\tablecaption{Diffuse Emission Line Fluxes\label{table-flux}}
\tablehead{
& \colhead{Observed Flux\tablenotemark{a}} & Extinction-corrected Flux\tablenotemark{b} \\
\colhead{Line} & \colhead{($10^{-12}$ erg cm$^{-2}$ s$^{-1}$)} 
&  \colhead{($10^{-12}$ erg cm$^{-2}$ s$^{-1}$)} 
}
\startdata
\sili\ 1.645 $\mu$m  &  0.056 (0.0074) & 0.30 (0.04) \\
\siv\ 10.51 $\mu$m  & 4.31   & 7.92 \\
\siii\  18.71 $\mu$m  &  2.26 & 3.19  \\
\oiv+\feii\  25.91 $\mu$m  &  18.0 & 21.2  \\
\siii\    33.48 $\mu$m  &  5.65 & 6.37  \\
\silii\  34.81 $\mu$m  &  24.5 & 27.5  \\ 
\enddata
\tablenotetext{a}{Flux is the total flux between $v=-2600$ and +2600~\kms\ 
within the \spitzer\ IRS SH area (see Figure~\ref{fig-spitzer}).
Statistical uncertainties in the \spitzer\ MIR line fluxes are less than 1\%, while the 
absolute flux calibration uncertainties (1$\sigma$) are 20\% \citep{decin04}.} 
\tablenotetext{b}{The extinction correction has been made 
assuming $N_{\rm H}=1.9\times 10^{22}$~cm$^{-2}$ (or $A_V$=10 mag) which is the mean absorbing H column density toward the IRS SH area from X-ray analysis \citep{hwanglaming}.}
\end{deluxetable}

\end{document}